\documentclass[11pt,a4paper]{article}

\usepackage{jheppub}
\usepackage{amsfonts}
\usepackage{amssymb}
\usepackage{comment}
\usepackage{epsfig}
\usepackage{graphicx}
\usepackage{amsmath}
\usepackage{tabu}
\usepackage[utf8]{inputenc}
\usepackage{latexsym}
\usepackage{tabularx}
\usepackage[nottoc]{tocbibind}
\usepackage{subfigure}
\usepackage{multirow}
\usepackage{booktabs}
\usepackage{enumerate}
\usepackage{mathrsfs}
\usepackage{latexsym} 

\usepackage{epstopdf}
\usepackage[section]{placeins}
\usepackage{placeins}

\newcommand{\bea}{\begin{eqnarray}}
\newcommand{\eea}{\end{eqnarray}}
\newcommand{\bi}{\begin{itemize}}
\newcommand{\ei}{\end{itemize}}
\newcommand{\ben}{\begin{enumerate}}
\newcommand{\een}{\end{enumerate}}
\newcommand{\be}{\begin{equation}}
\newcommand{\ee}{\end{equation}}
\newcommand{\ba}{\begin{align}}
\newcommand{\ea}{\end{align}}
\newcommand{\comments}[1]{}

\newcommand\vo{{\mathcal{V}}}

\newcommand{\mc}{\mathcal}

\newcommand{\beqa}{\begin{eqnarray}}
\newcommand{\eeqa}{\end{eqnarray}}

\title{De Sitter from T-branes}

\author[1,2,3]{\small{Michele Cicoli,}}
\author[3,4]{\small{Fernando Quevedo,}}
\author[5,6,3]{\small{Roberto Valandro}}

\affiliation[1]{Dipartimento di Fisica e Astronomia, Universit\`a di Bologna, \\ via Irnerio 46, 40126 Bologna, Italy}
\affiliation[2]{INFN, Sezione di Bologna, via Irnerio 46, 40126 Bologna, Italy}
\affiliation[3]{ICTP, Strada Costiera 11, 34151 Trieste, Italy}
\affiliation[4]{DAMTP, Centre for Mathematical Sciences, Wilberforce Road, Cambridge, CB3 0WA, UK}
\affiliation[5]{Dipartimento di Fisica, Universit\`a di Trieste, Strada Costiera 11, 34151 Trieste, Italy}
\affiliation[6]{INFN, Sezione di Trieste, Via Valerio 2, 34127 Trieste, Italy}

\emailAdd{mcicoli@ictp.it}
\emailAdd{f.quevedo@damtp.cam.ac.uk}
\emailAdd{roberto.valandro@ts.infn.it}

\abstract{Hidden sector D7-branes with non-zero gauge flux are a generic feature of type IIB compactifications. 
A non-vanishing Fayet-Iliopoulos term induced by non-zero gauge flux leads to a T-brane configuration. Expanding the D7-brane action around this T-brane background in the presence of three-form supersymmetry breaking fluxes, we obtain a positive definite contribution to the moduli scalar potential which can be used as an uplifting source for de Sitter vacua. In this way we provide a higher-dimensional understanding of known 4D mechanisms of de Sitter uplifting based on hidden sector F-terms which are non-zero because of D-term stabilisation.}

\preprint{DAMTP-2015-87 \\
\phantom{a} \hfill{}}


\begin{document}

\maketitle

\bigskip

\section{Introduction}

Chiral 4D string compactifications have several generic features, like extra dimensions, D-branes, background and gauge fluxes. In recent years global string models compatible with full closed string moduli stabilisation and a semi-realistic chiral visible sector have been constructed \cite{Cicoli:2011qg,Cicoli:2012vw, Cicoli:2013mpa,Cicoli:2013cha}. These models have to satisfy some global consistency constraints, like RR tadpole cancellation, Freed-Witten anomaly cancellation and discrete K-theory charge cancellation \cite{Blumenhagen:2008zz,Collinucci:2008sq}. Moreover the presence of chiral visible sector D-branes can reduce the effects that can be used to fix the moduli \cite{Blumenhagen:2007sm}. 

One of the brane configurations should host a visible sector that includes the Standard Model while other brane configurations give rise to hidden sectors. These hidden sectors are in general needed in order to guarantee the global consistency of the underlying compactification. Even though the coupling of these hidden sectors to the Standard Model is only gravitational, they may still play an important r\^ole to determine the physics of the visible sector by participating in the process of moduli stabilisation and by providing candidates for inflation, dark matter, dark radiation and dark energy.

The strongest global constraints come from the cancellation of D7- and D3-brane charges. The typical presence of O7-planes induces a D7-brane charge that has to be cancelled by some D7-branes which typically belong to the hidden sector, especially in models where the visible sector is realised on D3-branes at singularities \cite{Aldazabal:2000sa,Cicoli:2012vw, Cicoli:2013mpa,Cicoli:2013cha}. In order to cancel also the D3-brane charge in the presence of a large number of positively contributing three-form fluxes \cite{Giddings:2001yu}, the D7-branes should in general wrap a high degree divisor in the compact manifold. In Large Volume Scenario (LVS) models \cite{Balasubramanian:2005zx, Conlon:2005ki}, these cycles have very large volume. Moreover, in the models studied so far with small $h^{1,1}({\rm CY})$ \cite{Cicoli:2011qg,Cicoli:2012vw, Cicoli:2013mpa,Cicoli:2013cha}, this divisor has another peculiar feature: the pull-back of the harmonic $(1,1)$-forms from the Calabi-Yau to the D7-brane divisor has a 1D image, implying that any two-form of this kind is proportional to the pull-back of the K\"ahler form. A gauge flux on such a D7-brane has a remarkable effect: it induces a never-vanishing (inside the K\"ahler cone) Fayet-Iliopoulos (FI) term in the 4D effective action obtained by compactifying the D7-brane worldvolume theory. An image of the pull-back with dimension one is a sufficient condition to have a non-vanishing FI-term.  
For more generic cases, the FI term is a linear combination of several K\"ahler moduli and might vanish inside the K\"ahler cone; other competing effects in the moduli stabilisation process could however fix the K\"ahler moduli such that the FI term is non-zero.

In order to have a supersymmetric configuration at leading order, the D-term potential must vanish inducing a non-zero VEV for some charged matter fields if the corresponding FI-term is also non-vanishing. This 4D result, that has been studied in concrete examples in \cite{Cicoli:2012vw, Cicoli:2013mpa,Cicoli:2013cha}, can also be derived from an 8D perspective: %as we show in this paper: 
by solving the 8D equations of motion, a non-zero flux induces a non-zero VEV for the adjoint scalar $\Phi$ living on the D7-brane stack, so that $[\Phi, \Phi^\dagger]\neq 0$. This condition is solved by a T-brane background \cite{Donagi:2003hh,Cecotti:2010bp,Donagi:2011jy}. 

T-branes are particular brane configurations for which the non-Abelian Higgs field $\Phi$ describing D-brane deformations is not diagonalisable but takes an upper triangular form. This configuration generates a positive definite contribution to the scalar potential if non-supersymmetric three-form fluxes are switched on. The volume dependence of this new contribution can be computed explicitly, and so its influence on moduli stabilisation can be explicitly analysed. The presence of this uplifting terms  allows for dS vacua by tuning background three-form fluxes.

The paper is organised as follows. In Section \ref{4DSec} we present the 4D mechanism to generate a dS minimum from a hidden sector D7-brane. The flux on such a brane induces a moduli-dependent FI term in the effective field theory. Vanishing D-terms in turn induce a non-zero VEV for a hidden sector matter field that when substituted in the F-term scalar potential gives rise to a positive definite moduli-dependent term that plays the r\^ole of an uplifting term. The positivity of the F-term is understood as a soft mass term for the corresponding hidden sector matter field. We also study in detail the tuning of microscopic parameters needed to achieve a viable dS vacuum. In Section \ref{8DSec} we present instead the main result of this article: we show that from a higher dimensional point of view the 4D mechanism can be understood in terms of the presence of a T-brane. We consider supersymmetry breaking imaginary self-dual (ISD) three-form and gauge field fluxes in (hidden sector) D7-branes. The gauge fluxes lead to a T-brane configuration. We expand the D7 action and find a positive definite contribution to the scalar potential that precisely reproduces the uplifting term found from the 4D effective field theory point of view. This provides a higher dimensional implementation of this mechanism. A concrete example is based on the LVS scenario of moduli stabilisation in which three-form fluxes break supersymmetry. We present the equivalent analysis for the KKLT case \cite{Kachru:2003aw} in Appendix \ref{App} where the vacua, as expected, are only AdS.
Finally in Appendix \ref{FIdetails} we briefly review the derivation of the 4D D-term potential from the D7 brane action summarising some useful conventions used in the literature.

\section{4D point of view: dS from hidden matter F-terms}
\label{4DSec}

In this section we review a mechanism for obtaining dS vacua based on F-terms of hidden sector matter fields which are non-zero due to D-term stabilisation. 

\subsection{Hidden D7-branes and gauge fluxes}

Hidden sector D7-branes are a generic feature of globally consistent compact Calabi-Yau (CY) models because of D7-tadpole cancellation \cite{Cicoli:2011qg,Cicoli:2012vw,Cicoli:2013mpa,Cicoli:2013cha}. In these backgrounds, the orientifold involution typically generates O7-plane fixed points. These objects have a RR D7-brane charge that is measured by $-8[O7]$, where $[O7]$ is the homology class of the four-cycle wrapped by the O7-plane. The simplest way to cancel this charge is to place four D7-branes (plus their images) on top of the O7-plane locus.
The resulting D7-wordvolume gauge group is $SO(8)$. Here, without affecting our final results, we consider a simpler situation: we recombine the four D7-branes into a D7-brane wrapping the invariant divisor $D_{\rm h}$ in the homology class $4[O7]$ (where ${\rm h}$ stays for `hidden sector'). Its image will wrap the same divisor, creating an $SU(2)\cong USp(2)$ stack (the diagonal $U(1)$ is projected out by the orientifold action on the Chan-Paton factors). These branes can support a flux $\mc{F}$ along the Cartan generator of $SU(2)$:\footnote{The flux on one brane is equal in absolute value to the flux on the other brane but with opposite sign.}
\be
\mc{F} = (2\pi\alpha') F - \iota^* B\,,
\ee
where $\iota^* B$ is the pullback of the NS-NS $B$-field on $D_{\rm h}$. In order to cancel the Freed-Witten anomaly \cite{Minasian:1997mm,Freed:1999vc}, the gauge flux $(2\pi\alpha') F$ has to satisfy the following quantisation condition:
\be
(2\pi\alpha')F +\frac{c_1(D_{\rm h})}{2} \,\, \in \,\, H^2(D_{\rm h},\mathbb{Z}) \,.
\label{FD7FWquant}
\ee 
In the chosen configuration $D_{\rm h}$ is an even cycle, and so $c_1(D_{\rm h})$ is an even integral two-form (in a CY $c_1(D)=-[\hat{D}]$). This means that the gauge flux is integrally quantised and in principle can be set to zero (for a different situation with non-spin D7-brane divisor, i.e. $c_1(D_{\rm h})$ odd, the gauge flux cannot vanish because of \eqref{FD7FWquant}). On the other hand, $\mc{F}$ may be forced to be non-zero by the presence of a half-integrally quantised $B$-field. This is typically the case for compactifications with non-perturbative moduli stabilisation.\footnote{A non-perturbative superpotential $W_{\rm np}$ can be generated by D-branes wrapping four-cycles $D_{\rm np}$ that are typically non-spin. In order to have a non-zero $W_{\rm np}$, the flux $\mc{F}_{\rm np}$ on these branes should vanish. If $F_{\rm np}$ is half-integral, the $B$-field must be half-integral as well to have $\mc{F}_{\rm np}=0$. Hence the $B$-field is in general chosen in such a way to cancel the flux $\mc{F}_{\rm np}$ on $O(1)$ E3-instantons wrapped around different divisors since they contribute to $W_{\rm np}$ only if they are invariant under the orientifold involution.}
Therefore $\mc{F}$ is generically non-zero, causing the $SU(2)$ gauge group to break down to $U(1)$, where the $U(1)$ group is anomalous. This anomaly gets cancelled by the Green-Schwarz mechanism and the $U(1)$ gauge boson becomes massive by eating up a combination of an open and a closed string axion. 

The closed string modulus $T_j$ whose real part parameterises the Einstein-frame volume of the divisor $D_j$ in units of $\ell_s=2\pi\sqrt{\alpha'}$ gets a flux-dependent $U(1)$-charge of the form:
\be
q_{{\rm h}j} = \frac{1}{\ell_s^4}\int_{D_{\rm h}} \hat{D}_j \wedge \mc{F}\,,
\ee
where $\hat{D}_{\rm h}$ is the two-form Poincar\'e dual to the divisor $D_{\rm h}$ (here we take the extra-dimensional coordinates dimensionful). Moreover the gauge flux $\mc{F}$ yields a moduli-dependent Fayet-Iliopoulos (FI) term which looks like (see Appendix \ref{FIdetails} for more details):
\be
\frac{\xi_{\rm h}}{M_P^2} = \frac{e^{-\phi/2}}{4\pi\vo}\frac{1}{\ell_s^4}\int_{D_{\rm h}} J \wedge \mc{F}= \frac{1}{2\pi}\sum_j \frac{q_{{\rm h}j}}{2}\,\frac{t_j}{\vo}= - \frac{1}{2\pi}\sum_j q_{{\rm h}j}\frac{\partial K}{\partial T_j}\,,
\ee
where $\phi$ is the dilaton ($e^{\langle\phi\rangle}=g_s$), $\vo$ is the Einstein-frame CY volume in units of $\ell_s=M_s^{-1}$ where the string scale is related to the Planck scale as $M_s = g_s^{1/4} M_P/\sqrt{4\pi\vo}$, $t_j$ are two-cycle moduli, $J= t_j \hat{D}_j$ is the K\"ahler form expanded in a basis of $(1,1)$-forms $\hat{D}_j$ and $K/M_P^2=-2\ln\vo$ is the tree-level K\"ahler potential for the $T$-moduli. 

A non-zero $\mc{F}_{\rm h}$ induces also D7 matter fields $\phi_j=|\phi_j|\,e^{{\rm i}\theta_j}$ charged under the anomalous $U(1)$. 
These are in the symmetric representation of the $U(1)$ group, i.e. their charges are $q_{\phi_j}=\pm2$.
States with charges of both signs are typically generated for D7-branes wrapping large degree divisors, like in the cases under study.\footnote{See \cite{Braun:2015pza} for concrete computations in analogous situations.} 
The number of chiral fields with charge $+2$ is \cite{Blumenhagen:2006ci}:
\be
I_{U(1)} = \frac{2}{\ell_s^2}\int_{D_{\rm h}\cap D_{\rm h}} \mc{F}\,.
\ee
There are also neutral chiral fields which are counted by $h^{0,2}(D_{\rm h})$ and parameterise deformations of $D_{\rm h}$.
The resulting D-term potential takes the form \cite{Haack:2006cy,Cremades:2007ig,Cicoli:2011yh}:
\be
V_D = \frac{1}{2{\rm Re}(f_{\rm h})}\left(\sum_j q_{\phi_j} \phi_j \frac{\partial K}{\partial \phi_j}+ \frac{M_P^2}{2\pi}\sum_j q_{{\rm h}j} \frac{\partial K}{\partial T_j}\right)^2 =\frac{\pi}{{\rm Re}(T_{\rm h})}\left(\sum_j q_{\phi_j} \frac{|\phi_j|^2}{s}-\xi_{\rm h} \right)^2\,,
\label{VD}
\ee
since $f_{\rm h}=T_{\rm h}/(2\pi)$ and the K\"ahler metric $\tilde{K}$ for charged D7 matter fields depends just on the real part of the axio-dilaton ${\rm Re}(S)=s$ \cite{Aparicio:2008wh,Conlon:2006tj}:
\be
K \supset \tilde{K} \sum_j |\phi_j|^2= \sum_j \frac{|\phi_j|^2}{s}\,.
\ee

\subsection{Uplifting term from hidden matter F-terms}

Writing the K\"ahler moduli as $T_i = \tau_i+{\rm i}\psi_i$, the vanishing D-term condition reads:
\be
\sum_j \frac{q_{\phi_j}}{s} |\phi_j|^2 = \xi(\tau_i)\,,
\label{Drel}
\ee
and fixes the combination of $|\phi_j|$ and $\tau_i$ corresponding to the combination of $\theta_j$ and $\psi_i$ eaten up by the anomalous $U(1)$ which acquires a mass of the form \cite{Cicoli:2011yh}:
\be
M_{U(1)}^2 \simeq \frac{M_P^2}{{\rm Re}(f_{\rm D7})}\left(f_{\theta}^2 + f_\psi^2\right)\,,
\label{MU1}
\ee
where for simplicity we focused on the case with just one open and one closed string mode charged under the anomalous $U(1)$. The two terms in (\ref{MU1}) are proportional respectively to the open and closed string axion decay constants $f_{\theta}$ and $f_\psi$ which are given by:
\be
f_{\theta}^2 = |\phi|^2 \simeq \xi \simeq \frac{\partial K}{\partial \tau}M_P^2\qquad \text{and} 
\qquad f_\psi^2 \simeq \frac{\partial^2 K}{\partial \tau^2} M_P^2\simeq \xi^2\,.
\ee
Since $\xi/M_P^2 \simeq t/\vo \simeq 1/\tau \ll 1$ for cycles in the geometric regime, we realise that $f_\theta \gg f_\psi$, and so the combination of axions eaten up is mainly given by $\theta$. In turn, the relation (\ref{Drel}) fixes $|\phi|$ leaving $\tau$ as a flat direction. Consequently the mass of the anomalous $U(1)$ is of order the Kaluza-Klein scale:
\be
M_{U(1)} \simeq \frac{f_{\theta}}{\tau^{1/2}}M_P\simeq \frac{M_P}{\vo^{2/3}} \simeq \frac{M_s}{\vo^{1/6}} \simeq \frac{1}{\rm Vol^{1/6}}\,,
\ee
since via dimensional reduction the Planck scale $M_P$ is related to the string scale $M_s$ as $M_P \simeq M_s \vo^{1/2}$. 
By substituting the VEV:
\be
\frac{|\phi|^2}{s} = \frac{\xi}{q_\phi} = \frac{q_T}{q_\phi}\,\frac{t}{4\pi\vo} M_P^2\,,
\ee
into the F-term scalar potential for the matter field $\phi$, one obtains a moduli-dependent positive definite contribution to the total scalar potential which can be used as an uplifting term. In fact, the main contribution to the F-term potential for $\phi$ comes from supersymmetry breaking effects which generate scalar soft masses of the form:
\be
m_0^2 = m_{3/2}^2 - F^I F^{\bar{J}} \partial_I \partial_{\bar{J}}\ln\tilde{K}\,,
\label{m0}
\ee
where $m_{3/2}= e^{K/2}|W|$ is the gravitino mass expressed in terms of the 4D K\"ahler potential $K$ and superpotential $W$. Given that the K\"ahler metric for D7 matter fields $\tilde{K}$ depends just on the axio-dilaton $S$ which is fixed supersymmetrically by turning on three-form background fluxes $H_3$ and $F_3$, i.e. $F^S=0$, the soft masses for D7-matter fields from (\ref{m0}) are simply given by:
\be
m_0^2 = m_{3/2}^2 > 0\,.
\ee
This important relation ensures that the F-term scalar potential for the canonically normalised hidden matter field $\varphi=s^{-1/2}\phi$ is positive definite and can indeed play the r\^ole of the uplifting term:
\be
V_{\rm up} = m_0^2 |\phi|^2 = m_{3/2}^2 |\phi|^2 = \frac{c_{\rm up}}{\vo^{8/3}} M_P^4\,,
\label{Vup}
\ee
where (writing $6 \vo = k t^3$):
\be
c_{\rm up} = e^{K_{\rm cs}} \left(\frac{6}{k}\right)^{1/3} \frac{q_T}{q_\phi}\,\frac{|W|^2}{8\pi s} >0\,.
\ee
Note that the three-form fluxes $H_3$ and $F_3$, which fix the dilaton $S$ as well as the complex structure moduli, are also responsible for breaking supersymmetry by inducing non-zero F-terms for the K\"ahler moduli $T$. 

\subsection{Gauge fluxes and non-perturbative effects}

Let us show how the uplifting term (\ref{Vup}) can be successfully combined with the effects which fix the K\"ahler moduli $T$ to obtain in the end a viable dS vacuum. A crucial ingredient to freeze the $T$-moduli is the presence of a non-perturbative superpotential $W_{\rm np}$ generated by either gaugino condensation on D7-branes or E3-instantons. 

Note that a non-vanishing gauge flux $\mc{F}$ on $D_{\rm h}$ might induce chiral states between the hidden sector D7-stack and any E3-instanton (the situation is very similar for the gaugino condensation case) which transform in the fundamental representation of the hidden $SU(2)$ gauge group. If the E3-instanton wraps the divisor $D_{E3}$ and carries no gauge flux, the number of these E3 zero-modes is given by \cite{Blumenhagen:2006xt,Ibanez:2006da,Florea:2006si, Blumenhagen:2009qh}:
\be
I_{D7-E3}= \frac{1}{\ell_s^2}\int_{D_{\rm h}\cap D_{E3}} \mc{F} \,.
\ee
If present, these charged E3 zero-modes can prevent the contribution of the E3-instanton to the superpotential if their VEV is zero, since they appear as prefactors in $W_{\rm np}$ \cite{Blumenhagen:2007sm,Blumenhagen:2008zz}.\footnote{If the E3-instanton wraps the same four-cycle wrapped by the magnetised D7-brane, i.e. $D_{E3}=D_{\rm h}$, the presence of charged E3 zero-modes is almost unavoidable.} Let us consider the simplest case with just one charged matter field $\phi$ and a K\"ahler modulus $T$ with the following $U(1)$ transformations:
\be
\delta\phi = {\rm i} q_\phi \phi\qquad \text{and} \qquad \delta T= {\rm i} \frac{q_T}{2\pi}\,.
\ee
The corresponding non-perturbative superpotential takes the form (setting from now on $M_P=1$):
\be
W_{\rm np} = A \phi^n e^{-\frac{2\pi}{N} T}\,,
\ee
and its $U(1)$ transformation looks like:
\be
\delta W_{\rm np} = W_{\rm np} \left(n \frac{\delta\phi}{\phi} - \frac{2\pi}{N} \delta T\right) 
= {\rm i} W_{\rm np}  \left(n q_\phi - \frac{q_T}{N}  \right)\,.
\ee
Hence $W_{\rm np}$ is $U(1)$-invariant only if:
\be
n q_\phi = \frac{q_T}{N}\,.
\label{U1cond}
\ee 
This condition can be seen as a constraint that removes the tuning freedom to obtain a dS vacuum in models where $T$ is fixed by non-perturbative effects as in the KKLT case \cite{Kachru:2003aw}. On the other hand, if $T$ is fixed via perturbative corrections like the volume modulus in LVS models, the condition (\ref{U1cond}) does not impose any restriction on the possibility to obtain dS vacua. The problem to uplift KKLT-like vacua using F-terms of hidden matter fields which are non-zero due to D-term stabilisation, can be seen also by noting that (\ref{U1cond}) can be rewritten as:
\be
q_\phi \frac{\partial W}{\partial \phi}\phi + \frac{q_T}{2\pi} \frac{\partial W}{\partial T} = 0\,.
\label{r}
\ee
Recalling that $D_I W = W_I + W K_I$ and the expression (\ref{VD}) for the D-term potential $V_D = D^2/\left(2{\rm Re}(f_i)\right)$, (\ref{r}) becomes:
\be
q_\phi \phi \frac{D_\phi W}{W} + \frac{q_T}{2\pi} \frac{D_T W}{W} = \left(q_\phi \phi \frac{\partial K}{\partial \phi} + \frac{q_T}{2\pi} \frac{\partial K}{\partial T} \right)= D\,,
\label{D-F}
\ee
showing how the D-term is proportional to a combination of F-terms. Hence vanishing F-terms necessarily imply a vanishing D-term. KKLT vacua are characterised by the fact that $D_T W=0$, and so (\ref{D-F}) with $D=0$ implies $D_\phi W=0$, showing the impossibility to obtain a dS vacuum following this mechanism. On the other hand, if $D_T W\neq 0$ as in LVS models \cite{Balasubramanian:2005zx,Conlon:2005ki}, $D_\phi W$ can also be non-vanishing giving a viable uplifting term even if $D=0$ at leading order. We shall discuss the LVS case in some more detail in the next subsection and describe the KKLT case in Appendix \ref{App}.

\subsection{dS LVS vacua}

Let us now consider LVS vacua \cite{Balasubramanian:2005zx,Conlon:2005ki}\, with magnetised hidden sector D7-branes and E3-instantons where $K$ and $W$ take the form:
\be
K = -2\ln\left(\vo+\frac{\zeta s^{3/2}}{2}\right) + c\,\frac{\phi\bar{\phi}}{s}\qquad\text{and}\qquad W=W_0 + A_s e^{-a_s T_s}\,,
\ee
where $\zeta=-\frac{\chi(CY)\zeta(3)}{2(2\pi)^3}$ is a constant controlling the leading order $\alpha'$ correction and the CY volume depends on the big cycle $\tau_b$ and the small cycle $\tau_s$ as:
\be
\vo = \lambda_b\tau_b^{3/2}- \lambda_s \tau_s^{3/2}\,,
\ee
where $\lambda_b$ and $\lambda_s$ depend on the triple intersection numbers. The crucial difference with the KKLT case is that only $T_s$ is fixed by non-perturbative effects while $T_b$ is stabilised thanks to perturbative $\alpha'$ corrections to $K$ leading to a supersymmetry breaking AdS vacuum for $c=0$. Hence we can perturb this vacuum with the inclusion of magnetised branes and hidden matter fields which can lead to a dS solution. 

In order to generate a non-vanishing $T_s$-dependent contribution to $W$, the $B$-field is generically chosen to cancel the FW anomaly on the small cycle $T_s$. This leads to a non-vanishing gauge flux on the stack of hidden sector D7-branes wrapping the big cycle $T_b$ \cite{Cicoli:2012vw}. Thus $T_b$ acquires a non-zero $U(1)$-charge $q_b$ generating a moduli-dependent FI-term. The corresponding D-term potential reads:
\be
V_D = \frac{\pi}{\tau_b} \left(\frac{q_\phi}{s} |\phi|^2 - \xi_b\right)^2\,,
\label{dmatter}
\ee
where the FI-term takes the form:
\be
\xi_b = - \frac{q_b}{2\pi} \frac{\partial K}{\partial T_b} = \frac{3 q_b}{4\pi\tau_b} \,.
\ee
Therefore the total scalar potential is:
\be
V_{\rm tot} = V_D + V_F =  \frac{\pi}{\tau_b} \left(\frac{q_\phi}{s}  |\phi|^2 - \frac{3 q_b}{4\pi\tau_b}\right)^2
+ \frac{1}{s} m_{3/2}^2 |\phi|^2 +\frac{V_F(T)}{2 s}\,,
\label{Vtot}
\ee
where $m_{3/2}^2= e^K |W| \simeq W_0^2 /(2 s \vo^2)$ is the gravitino mass and $V_F(T)$ is given by \cite{Balasubramanian:2005zx, Conlon:2005ki}\footnote{Notice that in LVS, for $W_0\neq 0,$ the AdS minimum is already non-supersymmetric. The flux induced SUSY breaking is captured in the effective field theory by a non-vanishing F-term for the large modulus field $T_b$ which is proportional to $W_0$. Other fields such as the dilaton and $\phi$ itself will also contribute to the breaking of supersymmetry but their F-terms appear at higher orders in the $1/\vo$ expansion. The leading contribution to the goldstino comes from the fermionic component of $T_b$.}:
\be
V_F(T) = \frac{8}{3\lambda_s}(a_s A_s)^2 \sqrt{\tau_s} \,\frac{e^{- 2 a_s \tau_s}}{\vo}
- 4 a_s A_s W_0 \tau_s \frac{e^{-a_s \tau_s}}{\vo^2 } + \frac{3 \zeta s^{3/2} W_0^2}{4 \vo^3}\,.
\label{v0}
\ee
Note that the prefactor of the non-perturbative effect does not depend on the matter field $\phi$ since $T_s$ does not get charged under the anomalous $U(1)$. Minimising with respect to $\phi$ we find:
\be
\frac{q_\phi}{s} |\phi|^2 = \xi_b - \frac{m_{3/2}^2 \tau_b}{2\pi q_\phi}\,.
\label{minphi}
\ee
Substituting this VEV back in (\ref{Vtot}) we obtain (writing $\tau_b = \left(\vo/\lambda_b\right)^{2/3}$ and setting $s = g_s^{-1}$):
\be
V_{\rm tot} = \frac{g_s}{2} \left[c_{\rm up}\, \frac{W_0^2}{\vo^{8/3}}\left(1- \frac{c_{\rm sub}}{\vo^{2/3}}\right)+V_F(T)\right]\,,
\label{yt}
\ee
with:
\be
c_{\rm up} = \frac{3 q_b\lambda_b^{2/3}}{4\pi q_\phi}\qquad\text{and}\qquad c_{\rm sub} = \frac{W_0^2 g_s}{6 q_\phi q_b \lambda_b^{4/3}}\,.
\label{cup}
\ee
The first term in \eqref{yt} is the uplifting term (neglecting corrections proportional to $c_{\rm sub}$ which are subleading for $\vo\gg 1$):
\be
V_{\rm{up}}= \frac{g_s c_{\rm up}}{2} \frac{W_0^2}{\vo^{8/3}}\,.
\label{uplift4D}
\ee
This term modifies the scalar potential such that it now admits a dS minimum. In fact, minimising (\ref{yt}) with respect to $\tau_s$ we obtain:
\be
\vo = \frac{3 \lambda_s W_0\sqrt{\tau_s}}{4 a_s A_s}\, e^{a_s \tau_s}\,\frac{\left(1-\epsilon\right)}{\left(1-\epsilon/4\right)}\qquad\text{with}\qquad \epsilon \equiv\frac{1}{a_s \tau_s}\sim\mc{O}\left(\frac{1}{\ln\vo}\right)\ll 1\,.
\label{tsVEV}
\ee
Substituting this result back in (\ref{yt}) we find an effective potential for the volume $\vo$:
\be
V_{\rm tot}(\vo) = \frac{g_s W_0^2}{2 \vo^3} \left[c_{\rm up}\, \vo^{1/3}-\frac{3 \lambda_s}{2 a_s^{3/2}}\, \epsilon^{-3/2} \left(1-\left(\frac{3\epsilon}{4}\right)^2\right) + \frac{3 \zeta}{4 g_s^{3/2}}\right]\,,
\label{VFfinal}
\ee
where:
\be
\epsilon^{-1} = \ln\left( c_s\frac{\vo}{W_0}\right) -\ln \sqrt{\tau_s}+ \ln\frac{\left(1-\epsilon \right)}{\left(1-\epsilon/4 \right)} 
= \ln\left( c_s\frac{\vo}{W_0}\right)\left[ 1-\frac{\epsilon}{2}\ln \tau_s + \mc{O}(\epsilon^2)\right]\,,
\label{at}
\ee
with $c_s = 4 a_s A_s/(3 \lambda_s)$. Given that $\ln \tau_s\sim \mc{O}(1)$, (\ref{at}) scales as $\epsilon^{-1}= \ln\left( c_s\frac{\vo}{W_0}\right)\left(1+\mc{O}(\epsilon)\right)$, and so it can be rewritten iteratively as:
\be
\epsilon^{-1} = \ln\left( c_s\frac{\vo}{W_0}\right) -\frac 12 \ln \left[\frac{1}{a_s} \ln\left( c_s\frac{\vo}{W_0}\right)\right]+\mc{O}(\epsilon)\,.
\label{at2}
\ee
The uplifting term (\ref{VFfinal}) does not have a large tuning freedom since, as can be seen from (\ref{cup}), $c_{\rm up}$ is an $\mc{O}(1)$ number which depends just on triple intersection numbers and gauge flux quanta. On the other hand, the second and the third term in (\ref{VFfinal}) have a strong dependence on $g_s$ and $a_s = 2\pi/N$ (where $N$ is the rank of the condensing gauge group) while a milder logarithmic dependence on $W_0$. Hence a vanishing vacuum energy can be obtained by tuning the depth of the original AdS minimum by a proper choice of background fluxes, i.e. of $g_s$ and $W_0$. In general, for $N=1$ and large values of $W_0$ of order $100$ the scalar potential has a runaway behaviour unless $A_s$ is tuned to values of order $100$ to compensate the large value of $W_0$. In this case the potential has a dS minimum around $\vo\sim 10^6$ if $c_{\rm up}$ is of order $0.01$ and $g_s$ is appropriately tuned so that the third term in (\ref{VFfinal}) is of the same order of the first two. Larger values of $N$ and smaller values of $W_0$ give rise to dS minima for larger values of $\vo$. 

Let us see the tuning of the vacuum energy more in detail by minimising the potential (\ref{VFfinal}) with respect to $\vo$ that leads to:
\be
\frac{3\zeta }{4g_s^{3/2}} = \frac{3\lambda_s}{2}\, \tau_s^{3/2} \left[1-\frac{\epsilon}{2}+\mc{O}\left(\epsilon^2\right)\right]-\frac 89 \,c_{\rm up}\,\vo^{1/3}\,,
\label{VolVEV}
\ee
which substituted in (\ref{VFfinal}) gives a vacuum energy of the form
\be
\langle V_{\rm tot }\rangle = \frac{g_s}{18} \frac{W_0^2}{\vo^3} \left[c_{\rm up}\,\vo^{1/3}%\left(1+\mc{O}(\epsilon)\right)
-\frac{27\lambda_s}{4 a_s^{3/2}\sqrt{\epsilon}}\left(1-\frac98\epsilon +\mc{O}(\epsilon^2)\right)\right]\,,
\label{Vmin}
\ee
where $\vo$ (and hence $\epsilon$ given in \eqref{at2}) should be meant as a function of the underlying paramenters (like $W_0$ and $g_s$) following the minimizing equation \eqref{VolVEV}.
Setting this expression equal to zero and plugging the result for $c_{\rm up}$ back in (\ref{VolVEV}), we find:
\be
\tau_s= \left(\frac{\zeta}{2\lambda_s}\right)^{2/3}\frac{1}{g_s} \left[1+3\epsilon+\mc{O}\left(\epsilon^2\right)\right]\,.
\label{VolVEV2}
\ee
Comparing this result with the location of the AdS vacuum for $c_{\rm up}=0$ we find that the shift of the VEV of $\tau_s$ is proportional to $\epsilon$:
\be
\frac{\Delta \tau_s}{\left.\tau_s\right|_{c_{\rm up}=0}} = \frac{8\epsilon}{3} \left(1+\mc{O}(\epsilon)\right).
\ee
Plugging this result in (\ref{tsVEV}) the shift of the volume VEV is instead of order:
\be
\frac{\Delta \vo}{\left.\vo\right|_{c_{\rm up}=0}} = e^{8/3}\left(1+\mc{O}(\epsilon)\right)-1\simeq 14.4\left(1+\mc{O}(\epsilon)\right)-1\,,
\ee
showing that for $\epsilon\ll 1$, i.e. in the regime where higher order instanton contributions to $W$ can be safely neglected, the new dS vacuum is at values of the volume which are about one order of magnitude larger than those of the old AdS vacuum.

\begin{figure}[!ht]
\begin{center}
\includegraphics[width=0.5\textwidth, angle=0]{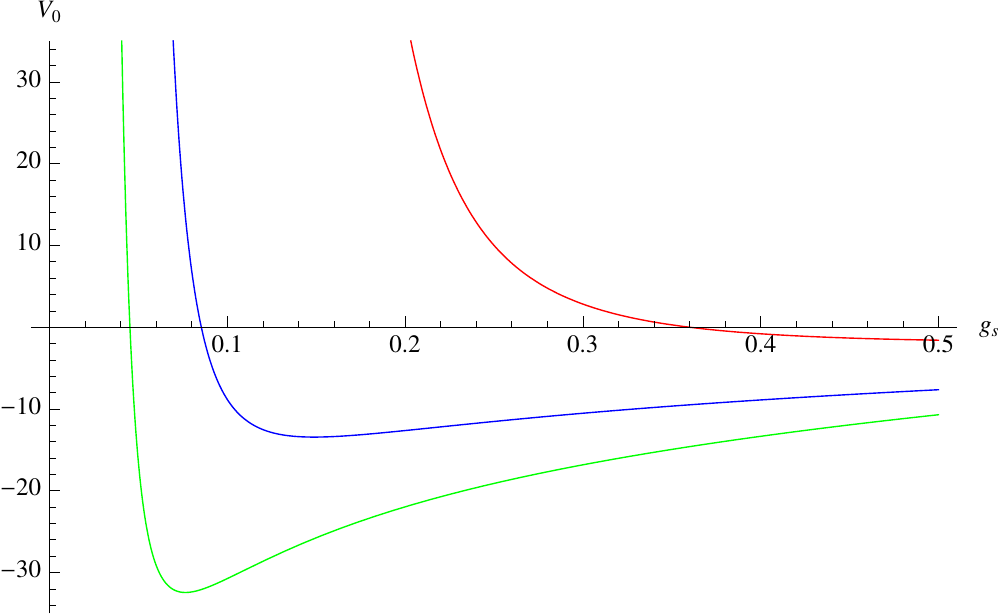}
\caption{Vacuum energy as a function of $g_s$ for $\lambda_s=\lambda_b=W_0=A_s=1$, $q_\phi=2 q_b$, $\zeta=2$ and $a_s=2\pi/N$. The intersection of the three curves with the abscissa shows the value of $g_s$ that gives a minimum where at leading order $\langle V\rangle=0$ for $N=2$ with $\vo \simeq 7.5\cdot 10^7$ (green), $N=6$ with $\vo\simeq 6.5\cdot 10^6$ (blue) and $N=10$ with $\vo\simeq 3\cdot 10^4$ (red).
Notice that we have plot $V_0\equiv \frac{18\vo^3}{g_sW_0^2} 
\langle V_{\rm tot }\rangle$ instead of $\langle V_{\rm tot }\rangle$, as we are interested in the zeros.} 
\label{Fig1}
\end{center}
\end{figure}

\begin{figure}[!ht]
\begin{center}
\includegraphics[width=0.5\textwidth, angle=0]{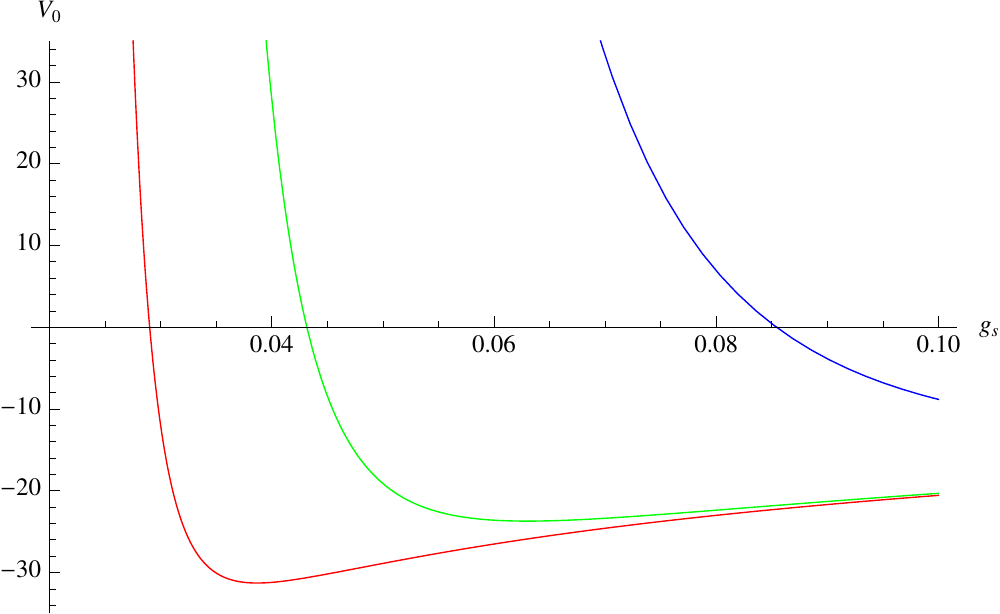}
\caption{Vacuum energy ($V_0$) as a function of $g_s$ for $\lambda_s=\lambda_b=A_s=1$, $q_\phi=2 q_b$, $\zeta=2$ and $a_s=2\pi/6$. The intersection of the three curves with the abscissa shows the value of $g_s$ that gives a minimum where at leading order $\langle V\rangle=0$ for $W_0=1$ with $\vo \simeq 6.5\cdot 10^6$ (blue), $W_0=10^{-5}$ with $\vo\simeq 1.8\cdot 10^7$ (green) and $W_0=10^{-10}$ with $\vo\simeq 3.5\cdot 10^7$ (red).} 
\label{Fig2}
\end{center}
\end{figure}

We conclude this section by showing in Figure \ref{Fig1} how the position of a leading order Minkowski minimum changes as a function of $g_s$ for three values of $N$, for fixed $W_0=1$ and natural values of the underlying parameters, while in Figure \ref{Fig2} we present the same behaviour for fixed $N=6$ and different values of $W_0$. To do this, we plot what is inside the square bracket in eq. \eqref{Vmin} and we call it $V_0$ (i.e. $V_0 \equiv \frac{18\vo^3}{g_sW_0^2} 
\langle V_{\rm tot }\rangle$).

\subsection{Open questions in the 4D approach}

The uplifting mechanism described above from the 4D effective point of view is based on two crucial ingredients, gauge and background fluxes, which induce non-zero hidden sector F-terms via D-term stabilisation. This mechanism for obtaining dS vacua seems to be rather generic and elegant but it raises some open questions which we list below:
\bi
\item Is it consistent to include an anomalous $U(1)$ in the 4D effective field theory if it acquires a mass of order the Kaluza-Klein scale?

\item What is the correct higher-dimensional understanding of this uplifting mechanism? Are we actually expanding the effective theory around a solution which at leading order is supersymmetric?

\item Does a non-zero VEV for $\phi$ correspond to a brane/image-brane recombination (or better a bound-state) that makes the whole procedure inconsistent since the flux $-\mc{F}$ on the image brane would cancel off with the flux $\mc{F}$ on the brane? This would mean that we have expanded around a supersymmetry breaking solution which might not be under control. 
\ei
These questions can be clearly answered only by developing a higher-dimensional understanding which we present in the following section.

\section{8D point of view: dS from T-branes}
\label{8DSec}

In this section we present an 8D description of the 4D dS vacua presented in Section \ref{4DSec}. The key-ingredient of this higher-dimensional understanding is the presence of T-branes induced by gauge fluxes.

\subsection{T-brane background due to gauge fluxes}

We consider a stack of $N$ D7-branes wrapping a compact four-cycle $D_{\rm h}$ in the CY three-fold $X_3$. Since $D_{\rm h}$ must be holomorphically embedded in $X_3$, it has an inherited K\"ahler form and it is a K\"ahler manifold. When the closed string fields have vanishing VEVs, the 8D theory living on their worldvolume is a twisted $\mc{N}=1$ SYM theory \cite{Beasley:2008dc}. The field content is made up of an 8D vector and a complex scalar $\Phi$ in the adjoint representation of the gauge group (in this case $U(N)$). 

The compactification on $D_{\rm h}$ breaks the 8D Poincar\'e invariance and breaks the $\mc{N}=1$ 8D supersymmetry. Thanks to the topological twist, four supercharges are still preserved. Hence the fields organise themselves in 4D $\mc{N}=1$ supermultiplets. We have a 4D vector multiplet with gauge field $A_\mu$, and two chiral multiplets with complex scalars $A$ and $\Phi$ in the adjoint representation. Due to the twist, the scalar $\Phi$ transforms in the canonical bundle of $D_{\rm h}$ like a $(2,0)$-form, while $A$ is a $(0,1)$-form on $D_{\rm h}$. The 8D supersymmetric equations of motion also split according to the breaking of the 8D Poincar\'e group. Along the compact direction they are given by:
\be
\label{Fterm7br}
\mc{F}^{0,2} =  \mc{F}^{2,0} = 0 \:, \qquad\qquad \bar{\partial}_A \Phi = \partial_A \Phi^\dagger = 0 \:,
\ee
where $\mc{F}$ is the curvature of the gauge bundle along $D_{\rm h}$ and $\partial_A$ is the holomorphic covariant differential relative to its connection and:
\be
\label{Dterm7br}
J_{({\rm h})}\wedge \mc{F} - \left[ \Phi,\Phi^\dagger \right] =0\:,
\ee
where $J$ is the K\"ahler form on $D_{\rm h}$ (in this case it is the pullback of the K\"ahler form of $X_3$, i.e $J_{({\rm h})}=\iota^* J$). Note that this relation is valid for a canonically normalised 8D scalar field. In 4D these equations boil down to F- and D-term conditions respectively. The second one will receive perturbative corrections. However in the cases we shall consider, the volume of $D_{\rm h}$ is very large and these corrections are negligible.

The structure group of the gauge bundle on $D_{\rm h}$ breaks the 4D gauge group to its commutant. As an illustrative example, we consider a stack of two D7-branes with gauge group $U(2)$ and we choose $\mc{F}$ to be an Abelian flux along the Cartan of $SU(2)$:
\be
 \mc{F} = f \otimes \left( \begin{array}{cc}
  1 & 0 \\ 0 & -1 \\
 \end{array}\right)\:,
\ee
where $f$ is a $(1,1)$-form on $D_{\rm h}$. We will consider the case when $f$ is the (non-trivial) pull-back of a two-form of $X_3$. This solves automatically the F-term eq. (\ref{Fterm7br}). 

Regarding the D-term eq. (\ref{Dterm7br}), the flux $\mc{F}$ and the K\"ahler form $J_{({\rm h})}$ determine if $\Phi$ can be zero or not. In particular, for simplicity (but without affecting our final results) we will restrict to the case when the pull-back map from two-forms on $X_3$ to two-forms on $D_{\rm h}$ has a 1D image (typical of the CY Swiss-cheese explicit models of \cite{Cicoli:2012vw, Cicoli:2013mpa,Cicoli:2013cha}) generated by the $(1,1)$-form $\omega$ on $D_{\rm h}$. This implies that for any combination $a_j \hat{D_j}$ its pull-back is $\iota^*a_j \hat{D}_j=a \, \omega$. Then, $J_{({\rm h})}=t \, \omega$ (where $t$ is a linear combination of the coefficient of the CY K\"ahler form $J=t_i\hat{D}_i$), $f= n \omega$ and
\be
J_{({\rm h})}\wedge f =(\, t \,n\,)\, \omega\wedge\omega \,.
\ee
This is different from zero for non-zero volume of $D_{\rm h}$ (in our cases this always happens if we want to keep the volume of $X_3$ different from zero as well, since vol$D_{\rm h} \propto t \sim \tau_b$).\footnote{
The expression $J_{({\rm h})}\wedge f $
is always different from zero under the sufficient assuption that the pull-back has a 1D image on $H^2(D_{\rm h})$ and that $D_{\rm h}$ is a large cycle. However it is a quite generic feature of a background with K\"ahler moduli stabilised by several effects whenever they imply $J_{({\rm h})}$ not orthogonal to the flux $f$.} 
Hence (\ref{Dterm7br}) implies that the commutator $[\Phi,\Phi^\dagger]$ is non-zero. The simplest solution is given by the nilpotent matrix:
\be
\label{TbrSol}
\Phi = \left( \begin{array}{cc}
0 & \varphi \\ 0 & 0 \\
\end{array}\right)\:,
\ee
that solves (\ref{Dterm7br}) if $\varphi\wedge \bar{\varphi}= t \,n\,\, \omega\wedge\omega$. Recall that $\varphi$ is a $(2,0)$-form. On a local patch of  $D_{\rm h}$ we have $\varphi=\hat{\varphi} \, dz_1\wedge dz_2$, where $\hat{\varphi}$ is a scalar field and $z_1$, $z_2$ are complex local coordinates. The solution (\ref{TbrSol}) is called a \emph{T-brane} (because of the triangular form of the matrix) \cite{Donagi:2003hh,Cecotti:2010bp,Donagi:2011jy}. This VEV for $\Phi$ does not change the D7-brane locus but anyway breaks the gauge group to the diagonal $U(1)$ of the original $U(2)$. %; in particular it breaks the Cartan $U(1)$, that disappears from the low energy spectrum. 
This is interpreted as a bound state of the two branes, even though the locus is made up of two copies of $D_{\rm h}$. In particular the broken (Cartan) $U(1)$ disappears from the low energy spectrum, where no FI term will appear.

\subsection{4D uplifting term}

As we will claim later, the solution described in the previous section is the leading piece of the solution in the presence of supersymmetry breaking ISD three-form fluxes for large volumes of $D_{\rm h}$ and $X_3$. We can then plug this into the 8D action of the D7-branes and expand around this solution.
When the supersymmetry breaking three-form fluxes are equal to zero, no dS uplift term is generated (by consistency). In fact the scalar potential at tree level is positive definite or identically zero (depending if we switch on supersymmetric fluxes). 

To see how the situation is modified by the presence of three-form fluxes, we need to compute the coupling of the relevant closed string $p$-form potential to the D7-worldvolume. Eventually we switch on a T-brane background and see how this interplays with three-form fluxes.

We start by reviewing what happens to a single D7-brane. Its 8D action is given by expanding the DBI and CS actions %\textbf{MC: Are you sure of $g_s$ factors? We should also be consistent and not use both $e^\phi$ and $g_s$}:
\be
\label{SDBICS}
 S = -\mu_7 \mbox{STr}\left\{ \int d^8\xi \,e^{-\phi} \sqrt{-\det(\iota^\ast e^{\phi/2}G_{AB}+  \mc{F}_{AB} )} - %g_s 
 \int (\iota^* C_6\wedge \mathcal{F} +\iota^\ast C_8) \right\}\,,
\ee
where $\mu_7^{-1}= (2\pi)^3(2\pi\alpha')^4$, %g_s$, 
$\iota^*$ is the pull-back map from the bulk to the D7-brane worldvolume and $\mc{F}=(2\pi\alpha') F-\iota^* B$. We want to understand what is the coupling of the bulk fields to the open string scalar $\Phi$. For a single D7-brane, this field controls the deformations of the D7-brane worldvolume in the orthogonal directions in the bulk. For a D7-brane we have one (complex) dimensional normal bundle. In local coordinates for an open patch of the CY three-fold $X_3$, $z^1$ and $z^2$ are complex coordinates tangent to the D7-brane and $z^3$ is orthogonal to it. We then make the identification $z^3=(2\pi\alpha') \Phi$ and expand the closed string fields around $z^3=0$, keeping only the leading terms and making the given identification. We will follow \cite{Camara:2004jj,Camara:2014tba} and refer to them for more details. 

The basic ingredients are three-form fluxes encoded into the complex three-form $G_3=F_3-\tau H_3$, where $F_3$ and $H_3$ are the field strengths of the (respectively) RR and NSNS two-form potentials and $\tau=e^{-\phi}+{\rm i} C_0$ is the axio-dilaton. The three-form $G_3$ can be decomposed into imaginary selfdual (ISD) and anti-selfdual (IASD) pieces:
\be
 G_3^\pm = \frac12(G_3 \mp i \ast_6 G_3)\,, \qquad \qquad \ast_6 G_3^\pm = \pm i\, G_3^\pm \:.
\ee
These components can be decomposed in irreducible pieces of the $SU(3)$ structure group of the CY manifold. For example the ISD component in ${\bf\overline{10}}$ representation of $SO(6)$ splits into ${\bf\overline{10}}={\bf\bar{1}}\oplus{\bf\bar{3}}\oplus{\bf\bar{6}}$ where the singlet is the $(0,3)$ piece $G_{\bar{k}\bar{j}\bar{\ell}}$ and the antisymmetric and symmetric representations are:
\be
 A_{\bar{i}\bar{j}} = \frac12 (\epsilon_{ik\ell}G_{\bar{k}\bar{\ell}j} - \epsilon_{jk\ell}G_{\bar{k}\bar{\ell}i} ) \:,  \qquad 
 S_{\bar{i}\bar{j}} = \frac12 (\epsilon_{\bar{i}\bar{k}\bar{\ell}}G_{\bar{j}k\ell} + \epsilon_{\bar{j}\bar{k}\bar{\ell}}G_{\bar{i}k\ell} )   \:.
\ee
For the IASD ${\bf 10}$ component the definitions are analogous (after taking the complex conjugate).
Following \cite{Camara:2004jj,Camara:2014tba} we neglect the non-primitive components $A_{\bar{i}\bar{j}}$ and $A_{ij}$ as they are incompatible with the CY topology (even though there can be a local component) and set to zero $S_{3i}$ and $S_{\bar{3}\bar{i}}$ since these components would generate a FW anomaly on the worldvolume of the D7-brane (that happens when $\iota^* G_3$ is a non-trivial three-form on the D7-brane worldvolume).

We can now expand the axio-dilaton and the $B$-field in powers of $\Phi$:
\bea
\tau &=& \frac{{\rm i}}{g_s}\left[ 1+\frac{\tau_{33}}{2}(2\pi\alpha')^2\Phi^2+\frac{\tau_{\bar{3}\bar{3}}}{2}(2\pi\alpha')^2\bar{\Phi}^2  +(2\pi\alpha')^2\tau_{3\bar{3}}|\Phi|^2 \right]\:, \\
B_{12} &=& -{\rm i}\,g_s\pi\alpha'\left((G_{\bar{1}\bar{2}\bar{3}})^\ast \Phi -\frac12 S_{\bar{3}\bar{3}}\bar{\Phi}-G_{123}\Phi+\frac12(S_{33})^\ast\bar{\Phi} \right) \:,\\
B_{1\bar{2}} &=& -{\rm i}\frac{g_s\pi\alpha'}{2}\left(-S_{\bar{2}\bar{2}} \Phi + (S_{\bar{1}\bar{1}})^\ast\bar{\Phi}-S_{11}\bar{\Phi}+(S_{22})^\ast\Phi   \right) \:,
\eea
where the $B$-field has been derived from $dB_2=-\frac{\mbox{Im}G_3}{\mbox{Im}\tau}$ and $\tau_{3\bar{3}}$ can be computed from the 10D supergravity equation of motion $\tau_{3\bar{3}} = \frac{1}{2{\rm i}} |G|^2$ (assuming that localised sources give a negligible contribution). 
Here and in the following we define (setting $A=0$): 
\be
 |G|^2 \equiv \frac{1}{12}  g^{mm'}g^{nn'}g^{kk'} G_{3|mnk}G_{3|m'n'k'} =  |G_{123}|^2 +\frac14\sum_{k=1}^3 |S_{kk}|^2 \,.
\ee
Finally, from:
\be
dC_6=H_3\wedge C_4 - \ast \mbox{Re}G_3 \qquad \text{and} \qquad dC_8=H_3\wedge C_6-\ast \mbox{Re}d\tau\,,
\ee
we obtain:
\bea
C_{\mu\nu\rho\sigma 12} &=& -{\rm i}\pi\alpha' \left( (G_{\bar{1}\bar{2}\bar{3}})^\ast \Phi    -G_{123}\Phi + \frac12 S_{\bar{3}\bar{3}}\bar{\Phi} +\frac12(S_{33})^\ast\bar{\Phi}  \right)\:,   \\
C_{\mu\nu\rho\sigma 1\bar{2}} &=& -{\rm i}\frac{\pi\alpha'}{2} \left( S_{\bar{2}\bar{2}} \Phi +(S_{22})^\ast\Phi - (S_{\bar{1}\bar{1}})^\ast\bar{\Phi} - S_{11}\bar{\Phi} \right) \:,   \\
C_{\mu\nu\rho\sigma 1\bar{1}2\bar{2}} &=& -\frac{g_s\pi^2\alpha'^2}{4} \left\{ \left[-2G_{123}S_{33} + (S_{11}S_{22})^\ast - S_{\bar{1}\bar{1}}S_{\bar{2}\bar{2}} + 2(G_{\bar{1}\bar{2}\bar{3}}S_{\bar{3}\bar{3}})^\ast \right] \Phi^2 + \text{h.c.} + \right. \nonumber \\
&& + \left[ |S_{33}|^2-|S_{\bar{3}\bar{3}}|^2 - 4|G_{\bar{1}\bar{2}\bar{3}}|^2 +  |S_{\bar{1}\bar{1}}|^2 + |S_{\bar{2}\bar{2}}|^2 + 4 |G_{123}|^2 - |S_{22}|^2 - |S_{11}|^2 \right] |\Phi|^2 \nonumber \\ 
&& \left. +{\rm i}\,g_s\pi\alpha' \left[\tau_{33} + (\tau_{\bar{3}\bar{3}})^\ast \right] \Phi^2 + \text{h.c.} \right\}\:.
\eea

We now plug these expressions in \eqref{SDBICS} and expand up to quadratic order, obtaining the following contribution to the 8D Lagrangian (we do not write the kinetic terms):
\bea
\label{8DLagrG3}
\mc{L} &\supset &
-2g_s| G |^2  |\Phi|^2  
  + \frac{g_s}{4} \left\{ (S_{\bar{3}\bar{3}})^\ast \left[ (G_{\bar{1}\bar{2}\bar{3}})^\ast - G_{123} \right]+ \frac12 (S_{11})^\ast \left[ (S_{22})^\ast - S_{\bar{2}\bar{2}} \right]  \right. \nonumber \\ 
	&& \left. - (G_{\bar{1}\bar{2}\bar{3}})^\ast S_{33} -  \frac12 (S_{22})^\ast  S_{\bar{1}\bar{1}} -2{\rm i}\tau_{33}  \right\} \Phi^2
	+ \text{h.c.} + [\, \Phi \, \partial A\,]\:,
\eea
where the terms in the last line are linear in $\Phi$ and in the derivatives of the connections.

This is the result for a single D7-brane. When we have several D7-branes on top of each other, each one will contribute with a term like \eqref{8DLagrG3}. Moreover, the gauge group is enhanced from $U(1)^N$ to $U(N)$, due to the massless strings stretching between different branes. The gauge field $A$ and the scalar field $\Phi$ live in the adjoint representation of $U(N)$. The terms in the Lagrangian that vanish for zero three-form fluxes are:
\bea
\label{8DLagrG3NA}
\mc{L} &\supset &
-2g_s| G |^2  \mbox{Tr}|\Phi|^2
  + \frac{g_s}{4} \left\{ (S_{\bar{3}\bar{3}})^\ast \left[ (G_{\bar{1}\bar{2}\bar{3}})^\ast - G_{123} \right]+ \frac12 (S_{11})^\ast \left[ (S_{22})^\ast - S_{\bar{2}\bar{2}} \right]  \right. \nonumber \\ 
	&& \left. - (G_{\bar{1}\bar{2}\bar{3}})^\ast S_{33} -  \frac12 (S_{22})^\ast  S_{\bar{1}\bar{1}} -2{\rm i}\tau_{33}  \right\} \mbox{Tr}\Phi^2
	+ \text{h.c.} + \mbox{Tr} (\Phi \, \partial A)\:,
\eea

We now want to insert in this Lagrangian the T-brane background (for simplicity we take $N=2$ D7-branes in the stack). Since the matrix $\Phi$ is nilpotent and $A$ is diagonal (it is along the Cartan), we have:
\be
{\rm Tr}\Phi^2 = 0 \qquad \qquad \mbox{and} \qquad \qquad {\rm Tr}(\Phi \, \partial A+ \Phi \, [A, A] )=0 \:.
\ee
The only contribution to the vacuum energy is then given by the first term in (\ref{8DLagrG3NA}). Hence the term contributing to the dS uplift is:
\bea
\mc{L}_{\rm up} &=& 2g_s| G |^2 \varphi\wedge\bar{\varphi}\:.
\eea

Let us see what is the volume dependence of this term. First of all, consider the three-form fluxes. 
Since they are quantised according to $\frac{1}{(2\pi)^2\alpha'}\int_{\Sigma_3} G_3 = n_F-\tau n_H$ (with $n_F,n_H\in \mathbb{Z}$), the field strength has to go like (using $M_s= \ell_s^{-1}$):
\be
G_{mnp} \sim (n_F-\tau n_H) M_s \sim  (n_F-\tau n_H) \frac{M_P}{\vo^{1/2}}\:.
\ee
Hence: 
\be
|G|^2 \sim (g^{-1})^3 M_s^2\sim \frac{1}{\vo}\frac{M_P^2}{\vo} = \frac{M_P^2}{\vo^2}\:.
\ee
Finally, from the T-brane equation of motion (\ref{Dterm7br}) and assuming that the D7-brane wraps a large cycle (whose volume is $\frac12\int_{D_{\rm h}} J_{({\rm h})}^2\simeq t^2\simeq \tau_b\simeq \vo^{2/3}$), we have:
\be
|\varphi|^2\sim \vo^{1/3}M_s^2 \sim \frac{M_P^2}{\vo^{2/3}} \:.
\ee
Hence the uplift term in the 4D potential takes the form:\footnote{No factor $\vo^{2/3}$ appears after integrating over $D_{\rm h}$. This is because we are taking a volume one surface and treating the volume modulus as a parameter. The canonical normalised 8D scalar that appears in (\ref{Dterm7br}) has already absorbed the factor coming from this modulus.}
\be
V_{\rm up} = \frac{C_{\rm up}}{\vo^{8/3}} M_P^4\:,
\label{VupliftFrom8D}
\ee
where $C_{\rm up} \sim 2 g_s n\, \int_{D_{\rm h}}  |\hat{G}|^2   \, \omega \wedge \omega$ is a constant that depends just on three-form and gauge fluxes (the dilaton is also a flux dependent function after stabilisation) since we set $\hat{G}=\frac{G}{M_P}\vo$.
This term is {\it positive definite} and is therefore suitable to uplift an AdS LVS vacuum to a dS solution. Note that (\ref{VupliftFrom8D}) has the same volume scaling as the uplifting term (\ref{uplift4D}) which we found using the 4D effective action.

Ref. \cite{Camara:2014tba} also studied corrections to the soft terms due to a non-zero gauge flux on the D7-brane worldvolume. In principle we should consider this effect as well, as the T-brane solution contains a non-zero flux. However, the extra contribution given by the presence of non-zero gauge fluxes is suppressed with respect to \eqref{VupliftFrom8D}. In fact, following \cite{Camara:2014tba}, we find a correction of the form $|G|^2|\Phi|^2 |\mc{F}|^2 M_s^{-4}$. The factor $|\mc{F}|^2 M_s^{-4}$ goes like $\vo^{-2/3}$ (due to two powers of the inverse metric) and provides a suppression factor of order $\vo^{-2/3}$ for the flux correction to the uplift potential.\footnote{As we have a non-abelian stack, one should have in principle included terms from the Myers action, proportional to Tr$ i_\Phi C$  and Tr$i_\Phi i_\Phi C$. For the solution \eqref{TbrSol} these are zero, as Tr$\Phi=0$ and Tr$[$Re$\Phi,$Im$\Phi]=0$.}

One could also question the procedure of expanding around the T-brane solution when three-form fluxes are switched on since they modify the equations of motion for $\Phi$. However a careful analysis reveals that the new solution is only a small perturbation around the used T-brane background and that this perturbation does not affect the uplift term.\footnote{Here we have considered the expansion of the 8D action in $\Phi$ up to quadratic order. This reproduces the controlled expansion in $\phi$ in the 4D approach, where it is more manifest that the higer order terms are suppressed by powers of the volume $\vo$.}

\section{Conclusions}

In this paper we have shown how a dS minimum can arise in type IIB CY compactifications if a T-brane is present among the hidden sector D7-branes.

We have first reviewed the 4D effective field theory picture that has been studied in~\cite{Cicoli:2012vw, Cicoli:2013mpa,Cicoli:2013cha}: the dS uplift term is generated by a fluxed D7-brane in the hidden sector. The flux on the D7-brane induces an FI-term. % that is always different from zero if the second homology of the divisor $D_i$ wrapped by the brane is of a particular (but non-rare) type, i.e. the image of the pull-back map from $H^{1,1}({\rm CY})$ to $H^{1,1}(D_i)$ is 1D. Then 
When this is different from zero, the D-term potential forces some charged hidden sector modes $\phi$ to get a non-zero VEV. When three-form fluxes are switched on, supersymmetry is broken and a positive soft supersymmetry-breaking mass-squared is produced for $\phi$. Since the VEV of $\phi$ is different from zero, due to the D-term condition, this mass term produces a constant positive definite term that can be used to achieve dS vacua. We have explained how this term can lead to a dS minimum by tuning $W_0$ and $g_s$.

This description can raise some concerns. First of all, in the effective field theory we have a non-vanishing FI-term. This means that the fluxed D7-brane is not supersymmetric. In fact, from the 4D point of view we see that it is the VEV of $\phi$ that restores supersymmetry (at the leading order). 
The field $\phi$ describes deformations of the D7-brane. The effective theory we have used is defined for the D7-brane on a given locus and corresponds to $\phi=0$. On the other hand, this effective theory is telling us that the VEV of $\phi$ is non-zero. This means that the right minimum is not the configuration around which we are expanding. Moreover, note that to cancel the FI term it is enough that only one charged hidden field takes a non-vanishing VEV. 

All these considerations led us to conjecture that the right supersymmetric configuration, that one should expand around, is a T-brane solution of the 8D equations of motion for the theory living on the D7-brane worldvolume. This takes place indeed when a non-zero gauge flux prevents the adjoint field $\Phi$ to commute with its conjugate. One can check that this is realised if an off-diagonal element of $\Phi$ gets a non-zero VEV that corresponds to a chiral state in the 4D effective field theory. To check this conjecture, we considered a simple $SU(2)$ model that has a T-brane solution, i.e. a proper VEV for $\Phi$. Three-form fluxes induce some particular couplings for $\Phi$ in the 8D worldvolume theory. After substituting the non-zero VEV of $\Phi$ in these couplings, we obtain exactly the uplift term found by using the 4D approach.

We have therefore succeeded to obtain metastable de Sitter string vacua starting from standard manifestly supersymmetric configurations with supersymmetry broken spontaneously. The mechanism is understood from both the 4D effective field theory and the higher dimensional one. The phenomenological implications of this class of de Sitter scenarios have been studied in \cite{Aparicio:2014wxa}. It would be interesting to construct a consistent compact setup with a global T-brane solution, as those constructed via tachyon condensation in \cite{Collinucci:2014qfa} and where moduli stabilisation is obtained in a dS LVS minimum. In principle this could be done in the global models presented in \cite{Cicoli:2012vw, Cicoli:2013mpa,Cicoli:2013cha} (where the uplift mechanism was studied only form the 4D point of view) by constructing the tachyon matrix realising the T-brane configuration and studying its stability and its low energy spectrum like in \cite{Collinucci:2014qfa}. We leave this for future work. %\textbf{MC: Why are our global models not good for T-branes? I thought that we are actually trying to say that they are examples of global T-branes.}

\section*{Acknowledgements}

We would like to thank A. Hebecker, C. Mayrhofer and T. Weigand for useful conversations.

\appendix

\section{AdS KKLT vacua}
\label{App}

In this appendix we consider for completeness the case of KKLT vacua. Unlike the LVS vacua discussed in the text in this case, as expected, there are only supersymmetric AdS vacua.
We focus on KKLT-like vacua with magnetised D7-branes and E3-instantons where $K$ and $W$ take the form:
\be
K = -3\ln(T+\overline{T}) + c\frac{\phi\bar{\phi}}{(T+\overline{T})^\alpha}\qquad\text{and}\qquad W=W_0 + A \phi^n e^{-a T}\,.
\ee
where $\alpha\geq 0$. The original KKLT model with just a closed string mode $T$ can easily be recovered by setting $c=n=0$. The two K\"ahler covariant derivatives read:
\bea
D_T W &=& - a A \phi^n e^{-a T} - \frac{3}{T+\overline{T}} \left(1+\frac{\alpha c}{3}\frac{\phi  \bar{\phi}}{(T+\overline{T})^\alpha}\right) W \nonumber \\
D_\phi W &=& n A \phi^{n-1} e^{-a T} + c\frac{\bar{\phi}}{(T+\overline{T})^\alpha} W
\eea
Imposing $D_T W =0$ implies:
\be
a A \phi^n e^{-a T} = - \frac{3}{T+\overline{T}} \left(1+\frac{\alpha c}{3}\frac{\phi  \bar{\phi}}{(T+\overline{T})^\alpha}\right) W \,.
\label{q}
\ee
Substituting this result into $D_\phi W$ we obtain:
\be
D_\phi W = \left[- \frac{3 n}{T+\overline{T}} + a c \frac{\phi\bar{\phi}}{(T+\overline{T})^\alpha}\left(1- \frac{n \alpha}{a(T+\overline{T})}\right)  \right] 
\frac{W}{a\phi}
\ee
$D_\phi W=0$ has a solution at:
\be
c \frac{\phi\bar{\phi}}{(T+\overline{T})^\alpha} = \frac{3 n}{a(T+\overline{T})}\left(1- \frac{n \alpha}{a(T+\overline{T})}\right)^{-1}\,.
\label{h}
\ee
Plugging this solution back in (\ref{q}) we find:
\bea
a A \phi^n e^{-a T} &=& - \frac{3}{T+\overline{T}} \left[1+\frac{n\alpha}{a(T+\overline{T})}\left(1- \frac{n \alpha}{a(T+\overline{T})}\right)^{-1}\right] W \nonumber \\
&=& - \frac{3}{T+\overline{T}} \left(1+\sum_{k=0}^\infty\epsilon^{k+1} \right) W\qquad\text{with}\quad \epsilon \equiv \frac{n\alpha}{a(T+\overline{T})}\ll 1\,.
\label{qnew}
\eea
The parameter $\epsilon$ is very small since $a (T+\overline{T})\gg 1$ in order to neglect higher order instanton contributions to $W$. It is clear that the solution with $n\neq 0$ is just a small perturbation with respect to the supersymmetric solution of the original KKLT model with $c=n=0$. Because of the relation (\ref{D-F}), the D-terms are automatically zero. Thus the new system keeps featuring the same supersymmetric solution of the original KKLT model. Let us study the form of the scalar potential to see if it can admit also dS solutions. Writing $T=\tau +{\rm i}\psi$ and $\phi=|\phi|\,e^{{\rm i}\theta}$ and minimising with respect to the axions, the F-term potential at leading order in an $\epsilon^{-1}$ expansion reads:
\be
V_F=\frac{a^2 A^2 |\phi|^{2 n} e^{-2 a \tau}}{6 \tau}-\frac{a A W_0 |\phi|^n e^{-a \tau}}{2 \tau^2}+c \frac{W_0^2}{(2\tau)^3}\frac{|\phi|^2}{(2\tau)^\alpha}\,.
\ee
On the other hand, the D-term potential at first order in $\epsilon^{-1}$ looks like:
\be
V_D = \frac{1}{\tau}\left(q_\phi c \frac{|\phi|^2}{(2\tau)^\alpha} - q_T \frac{3}{2\tau} \right)^2\,.
\ee
Minimising $V_{\rm tot}=V_D+V_F$ with respect to $|\phi|$ we find at leading order:
\be
\frac{\partial V_{\rm tot}}{\partial |\phi|} \simeq c q_\phi \frac{4}{\tau} \frac{|\phi|}{(2\tau)^\alpha} \left(q_\phi c\frac{|\phi|^2}{(2\tau)^\alpha} 
- q_T \frac{3}{2 \tau} \right).
\ee
Setting this to zero gives (we discard the solution $|\phi|=0$ since it gives a run-away for $\tau$ when substituted in $V_{\rm tot}$):
\be
q_\phi c\frac{|\phi|^2}{(2\tau)^\alpha} = q_T \frac{3}{2 \tau}\,,
\ee
which has the same leading order form of (\ref{h}) once we impose the $U(1)$-invariance condition (\ref{U1cond}).
Plugging this solution back in $V_{\rm tot}$ we end up with:
\be
V_{\rm tot}\simeq \frac{a^2 A^2 |\phi|^{2 n} e^{-2 a \tau}}{6 \tau}-\frac{a A W_0 |\phi|^n e^{-a \tau}}{2 \tau^2}+ \frac{q_T}{q_\phi} \frac{3 W_0^2}{16\tau^4}\,.
\ee
Minimising with respect to $\tau$ we find at leading order:
\be
\frac{\partial V_{\rm tot}}{\partial \tau}\simeq -\frac{a^3 A^2 |\phi|^{2 n} e^{-2 a \tau}}{3 \tau}+\frac{a^2 A W_0 |\phi|^n e^{-a \tau}}{2 \tau^2}=0\,.
\ee
This equation admits a solution at:
\be
a A |\phi|^n e^{-a \tau}= \frac{3 W_0}{2 \tau}\,,
\ee
which reproduces (\ref{qnew}) at leading order. Hence we conclude that this model can admit only supersymmetric AdS vacua.

\section{D-term potential from D7-branes}
\label{FIdetails}

In this appendix we shall work out the exact form of the gauge kinetic function and the FI-terms following \cite{Haack:2006cy} and appendix A.3 of \cite{Cicoli:2011yh}. The 4D D-term potential that arises from the dimensional reduction of the D7-brane action takes the form:
\be
V_D = \frac{\mu_7}{2}\,e^{-\phi} \frac{I_1^2}{I_2}\,,
\label{vd}
\ee
where:
\be
\mu_7 = (2\pi)^{-3} (2\pi\alpha')^{-4}\,,\qquad I_1=\int_{D_i} J\wedge \mc{F}\,,\qquad I_2 = \frac 12 \int_{D_i} J\wedge J\,.
\ee
Given that the real part of the gauge kinetic function is \cite{Haack:2006cy}:
\be
{\rm Re}(f_i) = g_i^{-2} = \mu_7 (2\pi\alpha')^2 e^{-\phi} I_2\,,
\label{gkf}
\ee
the D-term potential can be rewritten as:
\be
V_D = \frac{g^2}{2} \left( \mu_7 2\pi\alpha' e^{-\phi} I_1\right)^2 = \frac{g^2}{2}\,\xi_i^2\,.
\ee
Let us work out the expression of the FI-term $\xi$. If we expand $J$ and $\mc{F}$ in a basis of (1,1)-forms as $J = t_j^{(s)} \hat{D}_j$ ($t_j^{(s)}$ are two-cycle volumes in the string frame and in units of $\ell_s$) and $\mc{F} = f_k \hat{D}_k$, we get:
\be
I_1 = t_j^{(s)} f_k \int_{D_i} \hat{D}_j\wedge\hat{D}_k = t_j^{(s)} f_k \frac{1}{\ell_s^2}\int_{X_3} \hat{D}_j\wedge \hat{D}_j\wedge\hat{D}_k = k_{ijk} t_j^{(s)} f_k \ell_s^4\,,
\ee
where the triple intersection numbers $k_{ijk}$ are defined as:
\be
k_{ijk} = \frac{1}{\ell_s^6}\int_{X_3} \hat{D}_j\wedge \hat{D}_j\wedge\hat{D}_k \,.
\ee
Given that the gauge flux is quantised as ($\tilde{D}$ denotes the two-cycle dual to the divisor $D$):
\be
\int_{\tilde{D}_k} F = 2\pi n_k\qquad \Leftrightarrow\qquad\frac{1}{2\pi\alpha'}\int_{\tilde{D}_k} \mc{F} = 2\pi n_k\quad n_k\in \mathbb{Z}\,,
\ee
we find that the flux quanta $f_k$ take the form:
\be
\frac{f_j}{2\pi\alpha'} \int_{\tilde{D}_k} \hat{D}_j = 2\pi n_k\qquad\Rightarrow\qquad f_k = \frac{4\pi^2\alpha'}{\ell_s^2}\, n_k\,,
\ee
and so the FI-term becomes (using $\ell_s = M_s^{-1}$):
\be
\xi_i = \frac{e^{-\phi}}{(2\pi)^2}\left(\frac{\ell_s^2}{2\pi\alpha'}\right)^2 k_{ijk} t_j^{(s)} n_k\, M_s^2\,.
\label{xii}
\ee
From the dimensional reduction of the 10D action we have:
\be
M_s^2 = \frac{(2\pi)^3}{2}\, e^{\phi/2} \left(\frac{2\pi\alpha'}{\ell_s^2}\right)^4\frac{M_P^2}{\vo}\,,
\ee
where $\vo = e^{-3\phi/2} \vo_{(s)}$ is the CY volume in Einstein-frame. Hence (\ref{xii}) can be rewritten in Planck units as (where $t_j = e^{-\phi/2} t_j^{(s)}$):
\be
\frac{\xi_i}{M_P^2} =  \left(\frac{2\pi\alpha'}{\ell_s^2}\right) \frac{e^{-\phi/2}}{\ell_s^4} \frac{1}{2\vo}\int_{D_i} J\wedge \mc{F}=2\pi \left(\frac{2\pi\alpha'}{\ell_s^2}\right)^2 k_{ijk} n_k  \frac{t_j}{2 \vo}\,.
\label{xiFI}
\ee
The quantities $k_{ijk} n_k$ are integers and give the charge of the $j$-th K\"ahler modulus under the anomalous $U(1)$ living on the $i$-th divisor:
\be
q_{ij} = k_{ijk} n_k = \frac{1}{2\pi}\left(\frac{\ell_s^2}{2\pi\alpha'}\right)\frac{1}{\ell_s^4}\int_{D_i} \hat{D_j}\wedge \mc{F}\,.
\ee
Moreover, using four-cycle moduli:
\be
{\rm Re}(T_i) = \frac 12 \int_{D_i} J\wedge J = \frac 12 k_{ijk} t_j t_k\,,
\label{4T}
\ee
and the fact that $K/M_P^2 = -2\ln\vo$ we have that:
\be
\frac{\partial K}{\partial T_j} = - \frac{t_j}{2 \vo}\,,
\ee
and so (\ref{xiFI}) takes the form:
\be
\frac{\xi_i}{M_P^2} = - 2\pi \left(\frac{2\pi\alpha'}{\ell_s^2}\right)^2 q_{ij}\,  \frac{\partial K}{\partial T_j}\,.
\label{xiFin}
\ee
On the other hand, the gauge kinetic function (\ref{gkf}) can be rewritten in a more compact form using (\ref{4T}) as:
\be
{\rm Re}(f_i) = g_i^{-2} = \frac{e^{-\phi}}{(2\pi)^3 (2\pi\alpha')^2} \frac 12  \int_{D_i} J\wedge J = \frac{1}{(2\pi)^3}\left(\frac{\ell_s^2}{2\pi\alpha'}\right)^2 {\rm Re}(T_i)\,.
\label{gkfFin}
\ee
Therefore the exact expressions of $\xi_i$ and $f_i$ depend on the choice of the fundamental unit of lengths $\ell_s$. Let us give two examples:
\ben
\item For $\ell_s=\sqrt{\alpha'}$ we have \cite{Haack:2006cy}:
\be
f_k = (2\pi)^2 n_k\,, \,\,n_k\in\mathbb{Z}\qquad\text{and}\qquad M_s^2 = \frac{(2\pi)^7}{2}\,g_s^{1/2}\,\frac{M_P^2}{\vo}\,,
\ee
together with:
\be
f_i = \frac{T_i}{(2\pi)^5}\qquad\text{and}\qquad \frac{\xi_i}{M_P^2} = -(2\pi)^3 q_{ij}\,\frac{\partial K}{\partial T_j}\,.
\ee
This is the convention we followed in this paper. If the K\"ahler moduli are instead redefined as $\frac{T_i}{(2\pi)^5}\to T_i$ \cite{Haack:2006cy}, the new expression of $f_i$, $\xi_i$ and $M_s$ become:
\be
f_i = T_i\,,\qquad\frac{\xi_i}{M_P^2} = -\frac{q_{ij}}{(2\pi)^2}\,\frac{\partial K}{\partial T_j}\qquad\text{and}\qquad M_s^2 = \frac 12 \sqrt{\frac{g_s}{2\pi}}\,\frac{M_P^2}{\vo}\,.
\ee

\item For $\ell_s=2\pi\sqrt{\alpha'}$ we have \cite{Cicoli:2011yh}:
\be
f_k = n_k\,, \,\,n_k\in\mathbb{Z}\qquad\text{and}\qquad M_s^2 = \frac{g_s^{1/2}}{4\pi}\,\frac{M_P^2}{\vo}\,,
\ee
together with:
\be
f_i = \frac{T_i}{2\pi}\qquad\text{and}\qquad \frac{\xi_i}{M_P^2} = -\frac{q_{ij}}{2\pi} \,\frac{\partial K}{\partial T_j}\,.
\ee
If the K\"ahler moduli are redefined as $\frac{T_i}{2\pi}\to T_i$ \cite{Cremades:2007ig}, the new expression of $f_i$, $\xi_i$ and $M_s$ become:
\be
f_i = T_i\,,\qquad\frac{\xi_i}{M_P^2} = -\frac{q_{ij}}{(2\pi)^2}\,\frac{\partial K}{\partial T_j}\qquad\text{and}\qquad M_s^2 = \frac{1}{2 (2\pi)^2} \sqrt{\frac{g_s}{2\pi}}\,\frac{M_P^2}{\vo}\,.
\ee
\een

\end{document}